\documentclass[11pt]{amsart}
\usepackage{amsmath,amssymb,amsfonts}

\newtheorem{theorem}{Theorem}

\newtheorem{corollary}{Corollary}

\def\C{{\mathbb C}}
\def\R{{\mathbb R}}

\def\res{{\mathrm{res}}}

\def\res{{\mathrm{res}\ }}

\begin{document}

\title[Spectral conservation laws for the Melnikov systems]
{Spectral conservation laws for periodic nonlinear equations of the
Melnikov type}
\thanks{The work was supported by RFBR (grants 05-01-01032 (P.G.G.) and 06-01-00094
(I.A.T.)). The first author (P.G.G.) was also supported by the
Program of Russian Academy of Sciences ``Mathematical methods in
nonlinear dynamics'' and the second author (I.A.T.) was supported by
Max Plank Instit\"ut f\"ur Mathematik in Bonn.}
\author{P.G. Grinevich}
\address{Landau Institute of Theoretical Physics, Kosygin street 2,
117940 Moscow, Russia} \email{pgg@landau.ac.ru.}
\author{I.A. Taimanov}
\address{Institute of Mathematics, 630090 Novosibirsk, Russia}
\email{taimanov@math.nsc.ru}
\date{}

\dedicatory{We dedicate this article to our teacher S.P. Novikov
\\ on the occasion of his 70th birthday}

 \maketitle

In the seminal paper \cite{Novikov1} in 1974 S.P. Novikov, in particular,
established that the spectral curve of the one-dimensional periodic
Schr\"odinger operator
$$
H = -\frac{d^2}{dx^2} + u(x)
$$
is preserved when the real-valued potential $u(x,t)$ evolves via the
Korteweg--de Vries (KdV) equation and that for finite-zone (finite
gap) potentials the classical conservation laws, i.e. the
Kruskal--Miura integrals, are described in terms of branch points
for this curve. The spectral curve $\Gamma$ is a hyperelliptic
$$
\lambda^2 = Q(E)
$$
where
$$
Q(E) = (E-E_0)\dots(E-E_{2N})
$$
is a polynomial of degree $2N+1$ for $N$-zone potentials. It was
proved in \cite{Novikov1} that finite-zone potentials are exactly
solutions of the Novikov equations, i.e., stationary points of
higher KdV flows and their linear combinations, and that the KdV
flow on the set of $N$-zone potentials reduces to a completely
integrable finite-dimensional Hamiltonian system for which the ends
of the stability zones, i.e., $E_0,\dots, E_{2N}$, supply the
necessary family of first integrals.

The article \cite{Novikov1} was the starting point for the
development of the finite gap integration theory in which the
spectral curves play the main role.

In this article we consider the deformation of the spectral curve
via the periodic equations of the Melnikov type and we show that
although the spectral curve is not preserved it is deformed in such
a manner that it still gives many conservation laws for the system.

\medskip

{\bf 1. Introduction}

\medskip

We recall that the KdV equation
$$
u_t = 6uu_x - u_{xxx}
$$
has the Lax form
\begin{equation}
\label{lax}
H_t = [H,A]
\end{equation}
and as the {\it spectral curve} of $H$ parameterizes the Bloch
(--Floquet) functions which are formal eigenfunctions of $H$ (here
we do not mean that they lie in some nice functional space) and the
monodromy operator $\widehat{T}f(x) = f(x+T)$ where $T$ is the
period of $u(x)$:
\begin{equation}
\label{quasimomenta0}
H\psi = E\psi, \ \ \ \widehat{T}\psi(x) = e^{i\mu T}\psi(x)
\end{equation}
where $\mu$ is the quasimomentum which is defined on the spectral curve:
$\mu = \mu(\lambda,E)$.
The $t$-deformation of $u$ results in the deformation of $\psi$ via
the flow
$$
\psi_t = A\psi.
$$

Another form of soliton equations instead of the Lax form is the Manakov
triple:
\begin{equation}
\label{triple}
H_t = [H,A] + BH
\end{equation}
where $A$ and $B$ are differential operators. The main example is
given by the Novikov--Veselov (NV) equations \cite{NV2} for which
$H$ is a two-dimensional Schr\"odinger operator: $H =
\partial\bar{\partial}+u$. The {\it spectral curve of $H$ on the zero energy level} $\Gamma$
parameterizes only Floquet functions corresponding to  the zero
energy level:
\begin{equation}
\label{quasimomenta}
\begin{split}
H\psi = 0, \\
\psi(x+T_1,y) = e^{i\mu_1 T_1}\psi(x,y), \\
\psi(x,y+T_2) =
e^{i\mu_2 T_2} \psi(x,y)
\end{split}
\end{equation}
where $T_1$ and $T_2$ are the periods of $u$. This curve was first introduced
by Dubrovin, Krichever, and Novikov in \cite{DKN} where the inverse
problem at one energy level for two-dimensional Schr\"odinger
operators was posed and solved for
finite-zone operators (the spectral data for potential operators, i.e. with
no magnetic field, were later distinguished in \cite{NV1}).
Therewith the Floquet functions are deformed again via $\psi_t = A\psi$ and
hence the spectral curve is again preserved and may be considered itself as
a conservation law. Another equation of such triple form is
the modified Novikov--Veselov equation
for which $H$ is a two-dimensional Dirac operator and which being introduced
by Bogdanov found applications in the surface theory \cite{T1}.

Another generalizations of the Lax equations was proposed by Melnikov \cite{M1}
and later was also derived by Kuznetsov and Zakharov \cite{ZK}.
The general form of these equations is the following
extension of the Lax form:
$$
H_t = [H,A] + C
$$
where
$$
C = \sum_{n=1}^N C_n
$$
is the sum of differential operators $C_i$ with coefficients depending on
solutions $\phi_{i1},\dots,\phi_{ik_i}$ of the auxiliary linear problems
$$
H \psi_{ik} = \lambda_i \psi_{ik}, \ \ \ k=1,\dots,k_i.
$$
Very frequently these equations are called the equations with
self-consistent sources, each of them has a soliton predecessor of
the form $H_t=[H,A]$ and, for example, the KdV equation with
self-consistent force takes the form
\begin{equation}
\label{eq:KdV1}
u_t=\frac{1}{4}u_{xxx}-\frac{3}{2}uu_x+2\partial_x\sum\limits_{k=1}^N \psi_k(x,t) \psi^*_k(x,t)
\end{equation}
where $\psi_k(x,t)$, $\psi^*_k(x,t)$ are some solutions of the auxiliary
linear problem
\begin{equation*}
\begin{split}
(-\partial^2_x+u) \psi_k = E_k \psi_k, \\
(-\partial^2_x+u) \psi^*_k = E_k \psi^*_k.
\end{split}
\end{equation*}
To obtain a well-defined dynamics it is natural to assume that the
products of the eigenfunctions in (\ref{eq:KdV1}) are bounded.
The simplest choice for the periodic problem is the following: $\psi_k(x)$
is a Bloch eigenfunction and $\psi^*_k(x)$ is the Bloch eigenfunction
with the inverse Bloch multipliers.

The theory of such equations was developed in series of papers by Melnikov
\cite{M2,M3,M4,M5,M6} and others mostly for the case of functions fast
decaying at infinity.

In this article we show that in difference with soliton equations the spectral
curve is not preserved by these systems however it still gives many
conservation laws.

\medskip

{\bf 2. The spectral curve}

\medskip

The systems (\ref{quasimomenta0}) and (\ref{quasimomenta}) do not have
solutions for all possible values of constants, i.e. for all $E$ and $\mu$ in
the former case and for all $\mu_1,\mu_2$ in the latter case.
In fact, such solutions exist if and only if these constants satisfy
some analytical condition (``the dispersion laws''):
$$
F(E,\mu) = 0, \ \ \ G(\mu_1,\mu_2)=0.
$$
Each equation describes a complex curve $\Gamma \subset \C^2$ and to each
point of $\Gamma$ there corresponds a linear space of solutions to
the corresponding equation, (\ref{quasimomenta0}) or (\ref{quasimomenta}).
This picture was drawn in physical terms in \cite{Novikov2} and two different
methods for the justification of it were proposed by Krichever and the second
author (I.A.T.) (see \cite{Krichever2,T3,GT}).

Now to obtain the spectral curve $\Gamma_\psi$ we have to consider the
$\psi$-bundles formed by solutions to
(\ref{quasimomenta0}) or (\ref{quasimomenta}) and normalize
$\Gamma$ at such a manner that the pull-back of the $\psi$-bundle
onto $\Gamma_\psi$ under the projection
$$
\Gamma_\psi \to \Gamma
$$
form a bundle with fibers of constant dimension.
We refer for details to \cite{T3} and here demonstrate
this procedure by an important original example.

{\sc Example.} \cite{Novikov1} \ \
For the one-dimensional Schr\"odinger operator with a real-valued potential
the multipliers of
$\widehat{T}$ are defined on a Riemann surface (a complex curve) $\Gamma$
$$
\lambda^2 = \widehat{Q}(E)
$$
where $\widehat{Q}(E)$ is an entire function with infinitely many zeroes.
All zeroes lie on the real line.
To every point $P = (E,\lambda) \in \Gamma$ where $\widehat{Q}\neq 0$
there corresponds a one-dimensional space of solutions to
(\ref{quasimomenta0}). Let $E^\prime \in \R$ satisfy the following
conditions
\begin{enumerate}
\item
$\widehat{Q}$ has a zero at $E^\prime$ of
multiplicity two;

\item
to the point $(E^\prime,0)$ there corresponds a
two-dimensional space of solutions to (\ref{quasimomenta0}). (This, in
particular, implies that this is a double point on $\Gamma$.)
\end{enumerate}

\noindent
Let us unglue this double point and obtain another Riemann surface
$\Gamma^\prime$.
Then the $\psi$-bundle over $\Gamma$ is pulled back to a bundle
$\psi^\prime$ over $\Gamma^\prime$ with one-dimensional fibers
at the preimages of $(E^\prime,0)$. Moreover this bundle is holomorphic
near these points. We have

\begin{itemize}
\item
if all zeroes of $\widehat{Q}$ except finitely many satisfy conditions 1 and
2 above, then after ungluing all corresponding double points we obtain
a Riemann surface $\Gamma_\psi$ of finite genus and the one-dimensional
$\psi$-bundle over it. The surface $\Gamma_\psi$ is
defined by the equation
$$
\lambda^2 = Q(E)
$$
where $Q$ is a polynomial of odd degree, say $2N+1$.

It is said that this operator is {\it finite-zone} (or {\it finite gap}), and
it all zeroes of $Q$ are simple it is said that it has $N$ zones (gaps).
There is a function $\psi(P,x)$ meromorphic in $P \in \Gamma_\psi$
with the following asymptotic
\begin{equation}
\label{asymptotics}
\psi \approx e^{i\sqrt{E}x} \ \ \ \mbox{as $E \to \infty$}.
\end{equation}

Therewith the complex curve $\Gamma_\psi$ is compactified to an algebraic curve
by adding the point $E = \infty$ and $\psi$ becomes a meromorphic function
on $\Gamma_\psi \ \setminus \{E = \infty\}$ with the essential singularity
(\ref{asymptotics}) at $E=\infty$.
\end{itemize}

Here we remark that $\Gamma_\psi$ itself may have singularities and, in fact,
there is a tower of projections
$$
\Gamma_{\mathrm{norm}} \to \Gamma_\psi \to \Gamma
$$
where $\Gamma_{\mathrm{norm}}$ is the normalization of $\Gamma$.
The {\it multiplier mapping}
which corresponds to a point the set of ``multipliers'':
$$
{\mathcal M}: \Gamma \to \C^2, \ \ \ \ {\mathcal M}(E,\lambda) =
(E,\mu)
$$
is naturally ascends to this tower.

Above we explain how the spectral curve arises from the spectral theory of
differential operators.

However the strongest method for constructing exact periodic (and
also quasi-periodic) solutions of solitons equation, i.e. the {\it
Baker--Akhiezer function method} \cite{Krichever0}, starts with an
introduction of an algebraic curve $\Gamma$ and of a function $\psi$
(which may be a vector or even matrix function) with asymptotics of
the kind of (\ref{asymptotics}) at several points of $\Gamma$. It is
assumed that $\psi$ is defined by some additional data uniquely. The
function $\psi$ is a formal eigenfunction of some operator $H$ which
is uniquely reconstructed in terms of of algebraic functions
corresponding to $\Gamma$ from $\psi$. Such a function $\psi$ is
called the {\it Baker--Akhiezer function} of $H$, and the soliton
dynamics (\ref{lax}) or (\ref{triple}) is linearized in terms some
data coming in the definition of $\psi$ and this leads to explicit
algebra-geometrical formulas for so-called finite gap solutions of
soliton equations. The spectral curve $\Gamma$ is preserved by the
flow, i.e. the flow is {\it isospectral}.

Therewith for operators $H$ with periodic coefficients and with a nice
spectral theory (i.e. for which the existence of the dispersion laws may be
established) $\Gamma = \Gamma_\psi$ and $\psi$ is the section of the
$\psi$-bundle.

In this article we show that

\begin{itemize}
\item
in contrast with soliton equations the periodic equations of the
Melnikov type may be {\it almost isospectral}, i.e. it may preserve
${\mathcal M}(\Gamma_\psi)$ and deform $\Gamma_\psi$.
\end{itemize}

The first example of this effect was found by us in \cite{GT} and we
expose it in the next section.

\medskip

{\bf 3. The conformal flow for the (Weierstrass) potentials of tori in
$\R^3$ and $\R^4$}

\medskip

The author's interest to the study of the Melnikov-type equations was
partially motivated by the problem of conformal invariance of the
higher Willmore functionals.

By the generalized Weierstrass method, any torus in $\R^3$
is described in terms of the zero-eigenfunction $\psi$:
$$
{\mathcal D}\psi = 0,
$$
of a two-dimensional periodic operator
$$
{\mathcal D} =
\left(\begin{array}{cc} 0 & \partial \\
-\bar{\partial} & 0 \end{array}
\right) +
\left(\begin{array}{cc} U & 0 \\
0 & U \end{array}
\right)
$$
where the potential $U$ is real-valued
and any torus in $\R^3$ is described in terms of two solutions $\varphi, \psi$
to the equations
$$
{\mathcal D}\psi = 0, \ \ \ {\mathcal D}^\vee \varphi = 0
$$
where
$$
{\mathcal D} =
\left(\begin{array}{cc} 0 & \partial \\
-\bar{\partial} & 0 \end{array}
\right) +
\left(\begin{array}{cc} U & 0 \\
0 & \bar{U} \end{array} \right), \ \ \ {\mathcal D}^\vee =
\left(\begin{array}{cc} 0 & \partial \\
-\bar{\partial} & 0 \end{array}
\right) +
\left(\begin{array}{cc} \bar{U} & 0 \\
0 & U \end{array}
\right)
$$
are to conjugate periodic operators (see, for instance, \cite{T3}).
The spectral curve $\Gamma_\psi$ of ${\mathcal D}$ is naturally
defined (see (\ref{quasimomenta} and \S 2) and contains in itself
the information of the Willmore functional which is defined for all
closed surfaces immersed in $\R^4$ as follows
\begin{equation*}
{\mathcal W} (M) = \int_{M} |{\bf H}|^2 d\mu
\end{equation*}
where ${\bf H}$ is the mean curvature vector and $d\mu$ is the induced volume.

This functional is invariant with respect to conformal transformations of
the ambient space, i.e. if we have a conformal transformation
$f: \bar{\R}^4 \to \bar{R}^4$ which maps a compact surface without boundary
$M$ into a compact surface, then
$$
{\mathcal W}(M) = {\mathcal W}(f(M)).
$$
This follows from the conformal invariance of the form $(|{\bf H}|^2-K)d\mu$
where $K$ is the Gaussian curvature and the Gauss--Bonnet theorem by which
$\int Kd \mu$ equals $2\pi \chi(M)$, i.e. the topological quantity.

The soliton local deformations of surfaces in $\R^3$ and $\R^4$ via
the modified Novikov--Veselov (mNV) equation and the Davey--Stewartson (DS)
equation were introduced by Konopelchenko \cite{Kon1,Kon2}.
It appears that they preserve the tori globally and therewith preserve
the Willmore functional as well as the spectral curve \cite{T1,T3}.
hence it is natural to treat higher conservation laws of these hierarchies
as higher Willmore functionals.

The conformal invariance of the Willmore functional led
the second author (I.A.T.) to the conjecture that these higher Willmore
functionals
and the spectral curve  for tori in $\R^3$
themselves are conformally invariant \cite{T2}.

It was rather soon established
by the first author (P.G.G.) and M.U. Schmidt \cite{GS}
who considered the conformal flow, i.e. the Melnikov type flow, induced
on the potential $U$ by continuous conformal transformations:
$$
U_\tau = |\psi_2|^2 - |\psi_1|^2
$$
where the torus is defined via the Weierstrass formulas by
$\psi=(\psi_1,\psi_2)^\top$.
Under this deformation
the $\psi$-function on the spectral curve evolves in such a manner that
the quasimomenta are preserved.

In \cite{GT} we analyzed carefully this situation for the more general
case of tori in $\R^4$. It appears that the conformal flow on $U$
which corresponds to the following generator of the conformal group
\begin{equation*}
\begin{split}
\partial_{\tau}x^1 = 2 x^1 x^3, \ \ \ \
\partial_{\tau}x^2 = 2 x^2 x^3,\\
\partial_{\tau}x^3 = (x^3)^2 -(x^1)^2 - (x^2)^2 - (x^4)^2, \\
\partial_{\tau}x^4 = 2 x^4 x^3
\end{split}
\end{equation*}
has the Melnikov form:
\begin{equation}
\label{conformal-flow}
\begin{split}
\partial_\tau U= \varphi_1\bar\psi_1-\bar\varphi_2\psi_2, \\
\partial_\tau \bar U= \bar\varphi_1\psi_1-\varphi_2\bar\psi_2
\end{split}
\end{equation}
where $\psi = (\psi_1,\psi_2)^\top$ and $\varphi=(\varphi_1,\varphi_2)^\top$
define a torus in $\R^4$ via the generalized Weierstrass formulas.
It appears to be isospectral in the sense that all multipliers are preserved.
However we knew about several explicitly computed examples of  the Weierstrass
representations of tori which are the Clifford torus in $S^3$:
$$
x_1^2 + x_2^2 = x_3^2 + x_4^2 = \frac{1}{2}
$$
and its stereographic projection into $\R^3$ \cite{T3}. In these cases
the spectral curves $\Gamma_\psi$ are different: the complex projective line
$\C P^1$ in the former case and $\C P^1$ with two pairs of points glued into
two double points. However both tori are connected by a continuous conformal
transformation of $\bar{\R}^4$. A detailed analysis led us to
the following conclusion:

\begin{itemize}
\item
the conformal flow (\ref{conformal-flow}), i.e. a particular case of
Melnikov deformations of periodic operators, is only {\it almost
isospectral}, i.e. preserve the multipliers --- the complex curve
${\mathcal M}(\Gamma_\psi)$ --- and deform the spectral curve
$\Gamma_\psi$. In this particular case the deformation of
$\Gamma_\psi$ consists in gluing and ungluing double points.

\item
since the higher integrals of the mNV and the DS hierarchies are
described in terms of ${\mathcal M}(\Gamma_\psi)$, these integrals
are preserved and give us {\it spectral conservation laws} of the
conformal flow.
\end{itemize}

\medskip

{\bf 4. The Baker--Akhiezer function and kernel and the
$(\psi,\psi^\ast)$-representation of equations}

\medskip

Let us recall the definition of the Baker--Akhiezer function for the
KP equation \cite{Krichever0}.

Let $\Gamma$ be a smooth Riemann surface of genus $g$ with the
following data:

\begin{enumerate}
\item a divisor of poles $D=\gamma_1+\ldots+\gamma_g$;
\item a distinguished point $P$ with a local parameter $z=1/\lambda$.
\end{enumerate}

The Baker--Akhiezer function $\psi(\gamma,\vec t)$ depends on the
spectral parameter $\gamma\in\Gamma$ and of infinite set of real
variables $x=t_1$, $y=t_2$, $t=t_3$, $t_4$, $t_5$, \ldots, $\vec
t=(x,y,t,t_4,t_5,\ldots)$.  To avoid analytic problems it is
convenient to assume that $\vec t$ has only finite number of nonzero
entries.

For generic $\vec t$ there exists an unique function of
$\gamma\in\Gamma$ such, that:
\begin{enumerate}
\item $\psi(\gamma,\vec t)$ is meromorphic in $\gamma$ outside $P$ with
  simple poles at $\gamma_1$,\ldots,$\gamma_g$.
\item $\psi(\gamma,\vec t)=\exp\left[\sum\limits_{k>0} \lambda^k t_k\right]
\left(1+\sum\limits_{k>0}\frac{\chi_k(\vec t)}{\lambda^k}  \right)$
as $\gamma\sim P$.
\end{enumerate}
Let us define the potential $u(\vec t)$ by
\begin{equation}
\label{eq:potent}
u(\vec t) = 2 \partial_x \chi_1(\vec t).
\end{equation}
Then $u(\vec t)$ satisfy the KP hierarchy, and $\psi(\gamma,\vec t)$
is the common eigenfunction for all auxiliary linear problems. In
particular,
\begin{equation}
\label{eq:KP-auxiliary} -\psi_{xx}(\lambda,\vec
t)+\psi_{y}(\lambda,\vec t) + u(\vec t)\psi(\lambda,\vec t)=0.
\end{equation}
If $u$ is periodic in $x$ and $y$, then $\Gamma$ is the spectral
curve (on the zero energy level) of the operator $\partial_y -
\partial^2_x + u(x,y)$ \cite{Krichever2}.

Let us assume that $\Gamma$ is a hyperelliptic surface such that
$\lambda^2$ is a global meromorphic function  on $\Gamma$ with
exactly one second-order pole at $P$. Then
\begin{equation*}
\psi(\gamma,\vec t)=\exp\left[\sum\limits_{k>0} \lambda^{2k} t_{2k}\right]
\tilde\psi(\gamma,x,t,t_5,t_7,\ldots)
\end{equation*}
and $\tilde\psi(\gamma,\tilde t)$, $\tilde t
=(x,t,t_5,t_7,\ldots)$ is the Baker--Akhiezer function of the KdV
hierarchy \cite{DMN}. We shall omit the tilde sign in the KdV
formulas. In the KdV case we have
\begin{equation}
\label{eq:KdV-auxiliary}
-\psi_{xx}(\lambda,\vec t)+u(\vec t)\psi(\lambda,\vec t)=-\lambda^2\psi(\lambda,\vec t)
\end{equation}
instead of (\ref{eq:KP-auxiliary}).

In \cite{Cher} Cherednik has shown that all flows from the KdV
hierarchy are obtained as the expansion coefficients in
$\lambda^{-1}$  near $\lambda^{-1}=0$ for the following
$\lambda$-dependent nonlocal equation:
\begin{equation}
\label{eq:KdV3}
u_{\tau}= 2\partial_x (\psi_k(\lambda,x) \psi_k(-\lambda,x)).
\end{equation}
Here we assume that all times except $x$ are equal to 0.

\begin{theorem}
Let the source functions $\psi_k$ and $\psi^*_k$ in the right-hand
side of (\ref{eq:KdV1}) be the restrictions of the Baker--Akhiezer
function at some points of $\Gamma$:
\begin{equation*}
\psi_k=\psi(\lambda_k), \ \ \psi^*_k=\psi(-\lambda_k).
\end{equation*}
Then (\ref{eq:KdV1}) can be represented as the following linear
combination of the flows (\ref{eq:KdV3}):
\begin{equation*}
\begin{split}
u_{\tau} =  2\partial_x\left[-\res \raisebox{-3pt}
{$\Bigr|_{\gamma=P}$} \!\!\!\!\!\!\! ( \lambda^3
\psi(\lambda,x,\tau) \psi(-\lambda,x,\tau) d\lambda) + \right. \\
\left.
 + \sum\limits_{k=1}^N \psi(\lambda_k,x,\tau)
\psi(-\lambda_k,x,\tau) \right]
\end{split}
\end{equation*}
\end{theorem}

All the higher KdV flows are isospectral and form a commutative
algebra. Typically the complete algebra of symmetries for soliton
equations is non-commutative and contains both isospectral and
non-isospectral flows (see \cite{OS} for further references).

Orlov and Schulman suggested a generic approach for studying the
symmetry algebra based on the so-called infinitesimal dressing
\cite{OS}. It particular, in \cite{OS} it was shown that generators
$K_{mn}[u]$ of the algebra of all KdV and KP symmetries is obtained by
expanding the flow
\begin{equation}
\label{eq:KdV4} u_{\tau}= 2\partial_x (\psi_k(\lambda,\vec t)
\psi^*_k(\mu,\vec t))
\end{equation}
near the diagonal $\lambda=\mu$:
$$
2\partial_x (\psi_k(\lambda,\vec t)
\psi^*_k(\mu,\vec t))=\sum\limits_{m,n} K_{mn}[u] \left(\frac{1}{\lambda}\right)^{m}
\left(\frac{1}{\lambda}- \frac{1}{\mu} \right)^{n}
$$
at the point $P$ where $\lambda=\mu=\infty$ . For $n=0$ we have the
standard KP (KdV) hierarchy. The $n=1$ coefficients generate the
Virasoro algebra of non-sospectral symmetries, the $n>2$ symmetries
are not compatible with the KdV reduction.

 Here $\psi(\lambda,\vec t)$ is the
wave function for all auxiliary linear operators of the KP (KdV)
hierarchy, $\vec t=(x=t_1,t=t_3,t_5,\ldots)$ or $\vec
t=(x=t_1,y=t_2,t=t_3,t_4,\ldots)$ denotes the full set of KdV (KP)
times, and $\psi^*(\lambda,\vec t)$ satisfy the formal conjugate
linear problems. In the KdV case all auxiliary problems are
self-adjoint, therefore
$$
\psi^*(\lambda,\vec t)= \psi(-\lambda,\vec t).
$$

An arbitrary source function may be expanded in terms of
eigenfunctions products. Such expansions play a critical role in the
perturbations theory for soliton equations. The periodic
perturbation theory for $1$-dimensional finite-gap potentials and
for the 2-dimensional finite-gap at one energy potentials was
developed by Krichever \cite{Krichever1,Krichever2}. In particular,
he pointed out that it is natural to treat the conjugate
Baker--Akhiezer function $\psi^*(\lambda,\vec t)$ as a holomorphic
1-form in the spectral parameter $\gamma$ on $\Gamma\backslash P$.
It is defined by the following analytic properties:

\begin{enumerate}
\item $\psi^*(\gamma,\vec t)$ is an 1-form in $\gamma$, i.e.
in local coordinates it reads as $\psi^*(\lambda,\vec
t)=\tilde\psi^*(\lambda,\vec t)d\lambda$, where
$\tilde\psi^*(\lambda,\vec t)$ is an analytic function.
\item $\psi^*(\gamma,\vec t)$ is holomorphic in $\gamma$ outside $P$ with
  simple zeroes at $\gamma_1$,\ldots,$\gamma_g$.
\item $\psi^*(\gamma,\vec t)=\exp\left[-\sum\limits_{k>0} \lambda^k t_k\right]
(1+o(1))d\lambda$ as $\gamma\sim P$.
\end{enumerate}

The action of the Virasoro algebra symmetries on the finite-gap KP
solutions (these symmetries generically result in non-trivial
deformations of the complex structures on the spectral curves) was
studied by Orlov and the first author in \cite{GO}. In particular,
in \cite{GO} it was shown, that the infinitesimal deformations of
the Baker--Akhiezer function corresponding to the generators
(\ref{eq:KdV4}) (infinitesimal Darboux transformations of the
finite-gap KP solutions) are naturally written in terms of the
so-called Cauchy-Baker--Akhiezer kernel $\omega(\lambda,\mu,\vec
t)$:
$$
\delta\psi(\gamma,\vec t)=-\frac{\omega(\gamma,\mu,\vec t)}{d\mu}
\psi(\lambda,\vec t).
$$
The kernel $\omega(\lambda,\mu,\vec t)$ is defined by the following
analytic properties:
\begin{enumerate}
\item $\omega(\lambda,\mu,\vec t)$ is a meromorphic function in $\lambda$
  and a meromorphic 1-form in $\mu$ on $\Gamma\backslash P$.
\item For a fixed $\mu$ the function $\omega(\lambda,\mu,\vec t)$ has
  simple poles at the points $\mu$ $\gamma_1$,\ldots,$\gamma_g$.
\item For a fixed $\lambda$ the 1-form $\omega(\lambda,\mu,\vec t)$ has
 simple zeroes at $\gamma_1$,\ldots,$\gamma_g$ and a simple pole at
 $\lambda$.
\item $\omega(\lambda,\mu,\vec t)=\frac{d\mu}{\mu-\lambda} + O(1)$
near the diagonal $\lambda=\mu$.
\item For a fixed $\mu$ the function $\omega(\lambda,\mu,\vec t)
\lambda \exp\left[-\sum\limits_{k>0} \lambda^k t_k\right]$ is regular in $\lambda$ at the
point $\lambda=P$.
\item For a fixed $\lambda$ the 1-form $\omega(\lambda,\mu,\vec t)
\mu^{-1} \exp\left[\sum\limits_{k>0} \mu^k t_k\right]$ is regular in $\mu$
at the point $\mu=P$.
\end{enumerate}
For $\vec t = \vec 0$ this kernel coincides with the Cauchy kernel on
Riemann surfaces used by Koppelman \cite{Kopp}. For data generating
regular potentials $u(\vec t)$ the following explicit formula was
suggested in \cite{GO}:
\begin{equation*}
\begin{split}
\omega(\lambda,\mu,x,y,t_3,t_4,\ldots) =
\\
=
\int\limits_{x}^{\pm\infty} \psi(\lambda,x',y,t_3,t_4,\ldots)
\psi^*(\mu,x',y,t_3,t_4,\ldots)dx'.
\end{split}
\end{equation*}
The upper limit of the integral depends on the quasimomenta at the
points $\lambda$, $\mu$ and is chosen to make the integral convergent.
An analogous representation for the Cauchy kernels on Riemann
surfaces for systems with discrete $x$ was suggested earlier by
I.M.Krichever and S.P.Novikov in \cite{KN}.

\medskip

{\bf 5. The periodic Kadomstev--Petviashvili equation with a
self-consistent source}

\medskip

To integrate the KP equation with the self-consistent sources we
have to consider spectral curves with additional double points. Such
curves correspond to the solitons on the finite-gap background
\cite{Krichever3}. We assume that we have the same spectral data as
in the Section~4 plus $2N$ marked points $R^k_+$, $R^k_-$,
$k=1,\ldots,N$. Denote the local parameters near these points by
$\lambda$. The Baker--Akhiezer depends on $N$ extra real parameters
$\tau_1$, \ldots, $\tau_N$, $\vec\tau=(\tau_1,\ldots,\tau_N)$  and
has the following analytic properties:
\begin{enumerate}
\item $\psi(\gamma,\vec t,\vec\tau)$ is meromorphic in $\gamma$ outside $P$
  with $g+N$ simple poles at $\gamma_1$,\ldots,$\gamma_g$,
  $R^1_+$,\ldots,  $R^N_+$.
\item
\label{item2}
 $\res \raisebox{-3pt} {$\Bigr|_{\lambda=R^k_+}$}
\!\!\!\!\!\!\!\!\!\!\!\Psi(\lambda,\vec t,\vec\tau)d\lambda
= \tau_k \Psi(R^k_-,\vec t,\vec\tau)$.
\item $\psi(\gamma,\vec t,\vec\tau)=\exp\left[\sum\limits_{k>0} \lambda^k t_k\right]
\left(1+o(1)\right)$
as $\gamma\sim P$.
\end{enumerate}

The properties of the  conjugate Baker--Akhiezer 1-form are the
following:

\begin{enumerate}
\item $\psi^*(\gamma,\vec t,\vec\tau)$ is meromorphic in $\gamma$ outside $P$ with
  simple zeroes at $\gamma_1$,\ldots,$\gamma_g$ and simple poles at
  $R^1_-$,\ldots,  $R^N_-$.
\item
  $\res \raisebox{-3pt} {$\Bigr|_{\lambda=R^k_-}$}\!\!\!\!\!\!\!\!\!\!\!
  \Psi^*(\lambda,\vec t,\vec\tau)
= -\tau_k \Psi^*(\lambda,\vec t,\vec\tau)/d\lambda\raisebox{-3pt}
{$\Bigr| _{\lambda=R^k_+}$}$.
\item $\psi^*(\gamma,\vec t,\vec\tau)=\exp\left[-\sum\limits_{k>0} \lambda^k t_k\right]
(1+o(1))d\lambda$ as $\gamma\sim P$.
\end{enumerate}
The corresponding potential $u(\vec t,\vec\tau)$ is defined by the
formula (\ref{eq:potent})

\begin{theorem}
\label{th2}
Let $\Gamma$ be a Riemann surface of algebraic genus $g$
with the following KP data:
\begin{enumerate}
\item a divisor of poles $D=\gamma_1+\ldots+\gamma_g$;
\item a distinguished point $P$ with a local parameter
$z=1/\lambda$;
\item an additional collection of $2N$ points $R^k_+$, $R^k_-$, $k=1,\ldots,N$.
  Denote the local parameters near $R^k_+$, $R^k_-$ by $\lambda$.
\end{enumerate}
Then potential $u(\vec t,\vec\tau)$ defined above satisfy the
following equations with self-consistent sources:
\begin{equation}
\label{eq:tauder1}
\frac{\partial u(\vec t,\vec\tau)}{\partial\tau_k}= 2\partial_x
\frac{\psi(R^k_-,\vec t,\vec\tau) \psi^*(\lambda,\vec t,\vec\tau)}{d\lambda}
\biggr|_{\lambda=R^k_+}.
\end{equation}
\end{theorem}

{\sc Proof.} The Cauchy--Baker--Akhiezer kernel
$\omega(\lambda,\mu,\vec t,\vec\tau)$ corresponding to this spectral
data has the following analytic properties:
\begin{enumerate}
\item $\omega(\lambda,\mu,\vec t,\vec\tau)$ is a meromorphic function in $\lambda$
  and a meromorphic 1-form in $\mu$ on $\Gamma\backslash P$.
\item For a fixed $\mu$ the function $\omega(\lambda,\mu,\vec t,\vec\tau)$ has
  simple poles at the points $\mu$, $\gamma_1$,\ldots,$\gamma_g$,
  $R^1_+$,\ldots,$R^N_+$ .
\item \label{item3}$\res \raisebox{-3pt} {$\Bigr| _{\lambda=R^k_+}$}
\omega(\lambda,\mu,\vec t,\vec\tau)d\lambda = \tau
\omega(R^k_-,\mu,\vec t,\vec\tau)$.
\item For a fixed $\lambda$ the 1-form $\omega(\lambda,\mu,\vec t,\vec\tau)$
 has simple zeroes at $\gamma_1$,\ldots,$\gamma_g$ and simple poles at
 $\lambda$, $R^1_-$,\ldots, $R^N_-$.
\item $\res \raisebox{-3pt} {$\Bigr| _{\mu=R^k_-}$}
\omega(\lambda,\mu,\vec t,\vec\tau) = -\tau_k \omega(\lambda,\vec t,\vec\tau)/d\mu
\raisebox{-3pt} {$\Bigr| _{\mu=R^k_+}$}$.
\item \label{item6} $\omega(\lambda,\mu,\vec t,\vec\tau)=\frac{d\mu}{\mu-\lambda} + O(1)$
near the diagonal $\lambda=\mu$.
\item For a fixed $\mu$ the function $\omega(\lambda,\mu,\vec t,\vec\tau)
\lambda \exp\left[-\sum\limits_{k>0} \lambda^k t_k\right]$ is regular in $\lambda$ at the
point $\lambda=P$.
\item For a fixed $\lambda$ the 1-form $\omega(\lambda,\mu,\vec t,\vec\tau)
\mu^{-1} \exp\left[\sum\limits_{k>0} \mu^k t_k\right]$ is regular in $\mu$
at the point $\mu=P$.
\end{enumerate}
It is easy to check (see\cite{GO}):
\begin{eqnarray}
\label{eq:deromega}
& & \partial_x \omega(\lambda,\mu,\vec t,\vec\tau) =
-\psi(\lambda,\vec t,\tau) \psi^*(\mu,\vec t,\vec\tau), \\
& & \partial_y \omega(\lambda,\mu,\vec t,\vec\tau) =
-\psi_x(\lambda,\vec t,\vec\tau) \psi^*(\mu,\vec t,\vec\tau)+
\psi(\lambda,\vec t,\vec\tau) \psi^*_x(\mu,\vec t,\vec\tau).
\nonumber
\end{eqnarray}
We use the following formula:
\begin{equation}
\label{eq:dpsi}
\partial_{\tau_k} \psi(\lambda,\vec t,\vec\tau)= -
\frac{\omega(\lambda,\mu,\vec t,\vec\tau)}{d\mu} \biggr| _{\mu=R^k_+}
\!\!\!\!\!\!\!\!\cdot\psi(R^k_-,\vec t,\vec\tau).
\end{equation}
To prove (\ref{eq:dpsi}) is sufficient to check that the right-hand
side has the correct analytic properties. In particular, the
relation
$$
\partial_{\tau_k} \res \raisebox{-3pt} {$\Bigr| _{\lambda=R^k_+}$}
\psi(\lambda,\vec t,\vec\tau)d\lambda
= \tau_k \partial_{\tau_k} \psi(R^k_-,\vec t,\vec\tau)+
\psi(R^k_-,\vec t,\vec\tau)
$$
follows from the following expansion near the point $\lambda\sim R_+$:
$$
\frac{\omega(\lambda,\mu,\vec t,\vec\tau)}{d\mu} \biggr|_{\mu=R^k_+}=
\left[\tau \frac{\omega(R^k_-,\mu,\vec t,\vec\tau)}{d\mu} \biggr| _{\mu=R^k_+} -1\right]
\frac{1}{\lambda-R^k_+} +O(1).
$$
This expansion can be easily  derived from the properties
\ref{item3} and  \ref{item6} of the Cauchy--Baker--Akhiezer kernel.

Taking into account, that
\begin{equation}
\label{eq:auxiliary}
\psi_{xx}(\lambda)-\psi_{y}(\lambda)-u\psi(\lambda)=0, \ \
\psi^*_{xx}(\mu)+\psi^*_{y}(\mu)-u\psi^*(\mu)=0
\end{equation}
we obtain the following formula for the variation of $u=u(\vec
t,\vec\tau)$
\begin{equation}
\label{eq:deltau1}
u_{\tau_k} = \frac{\psi_{xx\tau_k}(\lambda)-\psi_{y\tau_k}(\lambda) -
u\psi_{\tau_k}(\lambda) }{\psi(\lambda)}.
\end{equation}
Substituting (\ref{eq:dpsi}), (\ref{eq:deromega}) into
(\ref{eq:deltau1}) we obtain (\ref{eq:tauder1}). We see, that the
right-hand side of (\ref{eq:deltau1}) turns out to be
$\lambda$-independent, therefore the deformation (\ref{eq:dpsi}) of
the Bloch function is admissible.
\begin{corollary}
\label{col:2}
Denote by $\widehat u(x,y,t)$, $\widehat\psi(\gamma,x,y,\tau)$,
$\widehat\psi^*(\gamma,x,y,\tau)$
the functions, obtained from $u(\vec t,\vec\tau)$, $\psi(\gamma,\vec t,\vec\tau)$,
$\psi^*(\gamma,\vec t,\vec\tau)$ by the following linear substitution:
\begin{eqnarray}
t_1=x, \ t_2=y,\ t_k = c_k \tau, \ k=3,\ldots,M, \ t_k=0, \ k>M, \\
\tau_k=\alpha_k+\beta_k \tau, \ k=1,\ldots,N.\nonumber
\end{eqnarray}
Then $\widehat u(x,y,t)$ solves the following Melnikov-type
equation:
\begin{equation}
\label{melnikov}
\begin{split}
\frac{\partial \widehat u(x,y,\tau)}{\partial\tau}=\\
 = \sum\limits_{k=3}^M c_k K_k[\widehat u] +2\partial_x
\sum\limits_{k=1}^N \beta_k \frac{\widehat\psi(R^k_-,x,y,\tau)
\widehat\psi^*(\lambda,x,y,\tau)}{d\lambda} \biggr|_{\lambda=R^k_+},
\end{split}
\end{equation}
where $K_k[\widehat u]$ denotes the $k$-th flow from the standard KP
hierarchy, the functions in the right-hand side satisfy
(\ref{eq:auxiliary})
\end{corollary}

{\sc Remark.} It is easy to notice from the
previous formulas that \footnote{The derivation of the similar fact
for the conformal flow (see Section 3) is exposed in \cite{GT}.}

{\sl if at $\tau=0$ the function $\widehat{u}$ is periodic in $x$
and $y$, $\psi$ is the Floquet eigenfunction of the operator $L =
\partial_y - \partial^2_x + u$, and for $k=1,\dots,N$ the
products $\widehat\psi(R^k_-,x,y,\tau)\widehat\psi^*(R^k_+,x,y,\tau)$
are periodic, then the evolution (\ref{melnikov}) preserves the
periodicity of $u$ and the multipliers of $\psi$.}

We conclude that

\begin{itemize}
\item
In fact, the Baker--Akhiezer function in Theorem \ref{th2} is
defined on the spectral curve with double points.  The double point
$(R^k_+,R^k_-)$ is unglued if and only if $\tau_k=0$. For generic
$\tau$ all double are present, i.e. , the spectral curve is obtained
from $\Gamma$ by pair-wise gluing points
$(R^1_+,R^1_-),\dots,(R^N_+,R^N_-)$, and the $k$-th double point is
unglued when $\alpha_k+\beta_k\tau=0$. If the initial spectral curve
is regular for $\tau=0$, then equations with self-consistent sources
immediately generate double points, which remains unglued for almost
all times.

Another example of such an effect is given by the conformal flow
(see Section 3).
\end{itemize}

This observation demonstrates the principal difference between the
Mel\-ni\-kov-type equations and the standard hierarchies like KdV,
nonliear Schr\"o\-din\-ger (NLS), sine-Gordon, or KP. The standard
hierarchies are isospectral, therefore the evolution can not
generate double points or unglue the existing double points in a
finite time.

In the case of the self-focusing NLS equation the spectral curves
with double points may correspond to regular space-periodic
solutions, associated with so-called whiskered tori. All double
points for such solutions remain glued for all values of $t$ but
become unglued in the limit $t\rightarrow\pm\infty$. In
\cite{AH1,AH2} it was shown that for a periodic solution
corresponding to a smooth curve the generation of double points
after arbitrary small perturbations results in numerical chaos.

\medskip

{\bf 6. An annihilation  of a soliton in a finite time}

\medskip

Let us discuss the simplest example -- the one-soliton solution of
the KdV equation.

The wave function $\psi(\lambda,x,c)$
of a one-dimensional Schr\"odinger operator:
$$
-\psi^{\prime\prime}+ u\psi = E\psi, \ \ \ \ u=u(x,c), \ \ E = (i\lambda)^2,
$$
has the
following form
$$
\psi(\lambda,x,c)=e^{\lambda x}\left(1+\frac{\chi(x,c)}{\lambda+\kappa}\right).
$$
The spectral curve $\Gamma$ is the Riemann sphere with a double
point: $\lambda=-\kappa$ and $\lambda=\kappa$ are glued together. We
assume that the divisor point is located at $\lambda=-\kappa$,
therefore we have:
$$
\left.\res\psi(\lambda,x,c)\vphantom{\int}\right|_{k=-\kappa}=-c
\psi(\kappa,x,c),
$$
(this relation coincides with the property (\ref{item2}) of the
Baker--Akhiezer function from the Section~5) and
$$
\chi(x,c)=\frac{-2c\kappa \ e^{\kappa x}}
{2\kappa e^{-\kappa x} + c e^{\kappa x}},
$$
By (\ref{eq:potent}) we obtain
\begin{equation}
\label{eq:1-sol} u(x,c) = 2\partial_x \chi(x,c)= \frac{-16c
\kappa^3} {\left(2\kappa e^{-\kappa x} + c e^{\kappa x}\right)^2}.
\end{equation}
The $c$-dynamics corresponds to the following choice of
Baker--Akhiezer function solutions:
$$
\psi_1(x,c)=\psi(\kappa,x,c)=
\frac{2 \kappa}{2\kappa e^{-\kappa x} + c e^{\kappa x}}.
$$
The conjugate Baker--Akhiezer function is defined by:
$$
\psi^*(\lambda,x,c)=\psi(-\lambda,x,c) dk.
$$
A simple straightforward calculation shows, that
\begin{equation}
\label{eq:1-sol2}
\partial_c u(x,c) = -2 \partial_x\psi^2(\kappa,x,c).
\end{equation}
Formula (\ref{eq:1-sol}) generates regular solitons for $c>0$,
singular solitons for $c<0$ and zero solution for $c=0$. If
$c\rightarrow 0$, the position of the soliton goes to the
$+\infty$.

The standard KdV evolution of a soliton is given by
(\ref{eq:1-sol2}) where $c=c(t)$ is governed by:
$$
\partial_t c(t) =\kappa^3 c(t).
$$

Let us consider the following Melnikov-type flow:
$$
u_t=\frac{1}{4}u_{xxx}-\frac{3}{2}uu_x+2\partial_x \psi^2(\kappa,t)
$$
The corresponding evolution of $c(t)$ is given by
$$
\partial_t c = \kappa^3 c - 1.
$$
We see that

\begin{itemize}
\item
starting with sufficiently small $c$ we reach the point $c=0$ at a finite
time. At this moment the double point on the spectral curve vanishes and the soliton annihilates.
\end{itemize}

By inverting the direction of evolution we obtain examples of
of such effects as

\begin{itemize}
\item
a creation of a soliton (there is no soliton at $c=0$ and it exists as soon as $c >0$);

\item
a capture of a soliton ($c \to \kappa^{-3}$ as $t \to \infty$),
\end{itemize}

\noindent
first observed by Melnikov \cite{M2}.

\end{document}